\documentclass[aps,prl,twocolumn,superscriptaddress,showpacs]{revtex4-1}

\usepackage{graphicx}
\usepackage{color}
\usepackage{amsbsy,amssymb,amsmath,amsfonts,bm}
\usepackage{float}

\begin{document}

\title{Magnetic field-tuned superconductor/insulator transition in TiN nanostrips}
\author{I. Schneider}
\affiliation{Institute of Experimental and Applied Physics, University of Regensburg, D-93040 Regensburg, Germany}
\author{K. Kronfeldner}
\affiliation{Institute of Experimental and Applied Physics, University of Regensburg, D-93040 Regensburg, Germany}
\author{T.\,I. Baturina}
\affiliation{Institute of Experimental and Applied Physics, University of Regensburg, D-93040 Regensburg, Germany}
\affiliation{Institute of Semiconductor Physics, 13 Lavrentjev Avenue, Novosibirsk, 630090 Russia}
\affiliation{Novosibirsk State University, Pirogova Str. 2, Novosibirsk 630090, Russia}
\author{C. Strunk}
\affiliation{Institute of Experimental and Applied Physics, University of Regensburg, D-93040 Regensburg, Germany}
\email[]{christoph-.strunk@physik.uni-regensburg.de}

\date{18. May 2018}

\begin{abstract}
We have measured the electric transport properties of TiN nanostrips with different widths. At zero magnetic field the temperature dependent resistance $R(T)$  saturates  at a finite resistance towards low temperatures, which results from quantum phase slips in the narrower strips. 
We find that the current-voltage ($I$-$V$) characteristics of the narrowest strips are equivalent to those of small Josephson junctions. Applying a transverse magnetic field drives the devices into a reentrant insulating phase, with $I$-$V$-characteristics dual to those in the superconducting regime. The results evidence that our critically disordered superconducting nanostrips behave like small self-organized random Josephson networks.
\end{abstract}



\maketitle

The loss of global phase coherence due to quantum phase slips (QPS) is a prime element in the understanding of resistive or even insulating states in one-dimensional (1d) superconducting systems \cite{Haviland2001,Arutyunov2008,Bezryadin2008}. 
In single Josephson junctions \cite{corlevi2006, Ergul2013} and  1d  Josephson junction arrays~\cite{Pop2010} phase slippage is meanwhile rather well understood. In small junctions, the Josephson coupling competes with Coulomb charging effects, leading to QPS. If the Josephson coupling energy $E_J$ is comparable to or smaller than the charging energy $E_C$, supercurrent is suppressed by QPS that persist down to zero temperature.  QPS can be either incoherent or coherent. The former are dissipative and leads to a finite resistance, whereas the latter give rise to the Coulomb blockade of Cooper pair tunneling, i.e., strongly insulating behavior below a critical voltage $V_c$. Above $V_c$ a typical back-bending ("Bloch-nose")  of the $I$-$V$-characteristics is observed \cite{corlevi2006} that indicates Bloch oscillations of the voltage. These are the electrodynamic dual to the current oscillations of the standard ac-Josephson effect \cite{likharev1985}. Achieving this dual to the Josephson effect in a circuit which can support a dc electric current is of fundamental interest for quantum metrology. 

The situation is less clear for homogeneously disordered nanowires. At higher temperatures, the phase slip rate, and thus the resistance, behaves thermally activated. In the zero temperature limit it either vanishes exponentially or saturates at a finite value by virtue of quantum tunneling. When the superconducting state in the wire is spatially homogeneous, the phase slips are delocalized along the wire, as opposed to an inhomogeneous superconductor, where local phase slips occur preferentially at weak spots.  To avoid inhomogeneities due to grain boundaries the wires are often made of amorphous materials like Mo$_\text{x}$Ge$_\text{1-x}$,  Nb$_\text{x}$Si$_\text{1-x}$ or Ti \cite{Bollinger2006,Bollinger2008,Webster2013,Lehtinen,Lehtinen2017}.  Coherent quantum phase slips were recently demonstrated in amorphous InO$_\text{x}$ wires \cite{Astafiev2012}.

For critically disordered films, i.e., films close to the disorder-induced superconductor-insulator transition (SIT), it has been suggested that the superconducting condensate is intrinsically fractionalized into small droplets \cite{Ovadyahu1994,Ghosal2001,Ghosal2001b}. 
 This prediction is meanwhile supported by accumulating experimental evidence in ultrathin films \cite{Baturina2007,Sacepe2008,Valles2009, Lemarie2013,Carbillet2016}. The experiments show that random fluctuations of the superconducting gap occur on a typical scale of 20-50\,nm, which is larger than the superconducting coherence length $\xi\simeq 9\,$nm. In TiN and NbN the gap fluctuations are uncorrelated with the underlying fine-crystalline structure of the metal film. 
 
 Hence, the question arises, how the disorder induced gap fluctuations manifest themself in narrow strips, when the strip width becomes comparable to the typical length scale of the fluctuations. We expect that the transport properties of the strips deviate from those of both two\-dimensional films and {homogeneous} nanowires in that random weak spots start to dominate their behavior. If indeed Josephson coupled \cite{footnote} granules are formed, the supercurrent will be dominated by the smallest Josephson coupling energy $E_J$ in the system. When $E_J$ is sufficiently small,  charging effects will reduce the maximal Josephson current via quantum fluctuations of the phase. In addition, we anticipate that the magnetic flux enclosed between adjacent granules will reduce the dominating Josephson coupling energy with respect to the charging energy $E_C$, resulting in Coulomb blockade and strongly insulating behavior.
 
Here, we address the above questions in a systematic study of the properties of nanostrips of width $8\xi<w<50\xi$ made of a critically disordered ultrathin superconducting TiN film of thickness $d<\xi$. 
We find a gradual suppression of the zero-resistance state with a low temperature resistance saturating at a value that increases as the strips narrow down.
The current-voltage characteristics strongly resemble those of small Josephson junctions for the narrowest strips. 
At finite magnetic field we observe reentrant behavior and highly insulating states; the $I$-$V$-characteristics being dual to those observed in the superconducting state in that current and voltage are interchanged. The insulating state is characterized by a critical voltage that is modulated by the magnetic field. Our results evidence that superconducting strips made from critically disordered films behave as random Josephson networks.

\begin{figure}[t]
\includegraphics[width=1\linewidth]{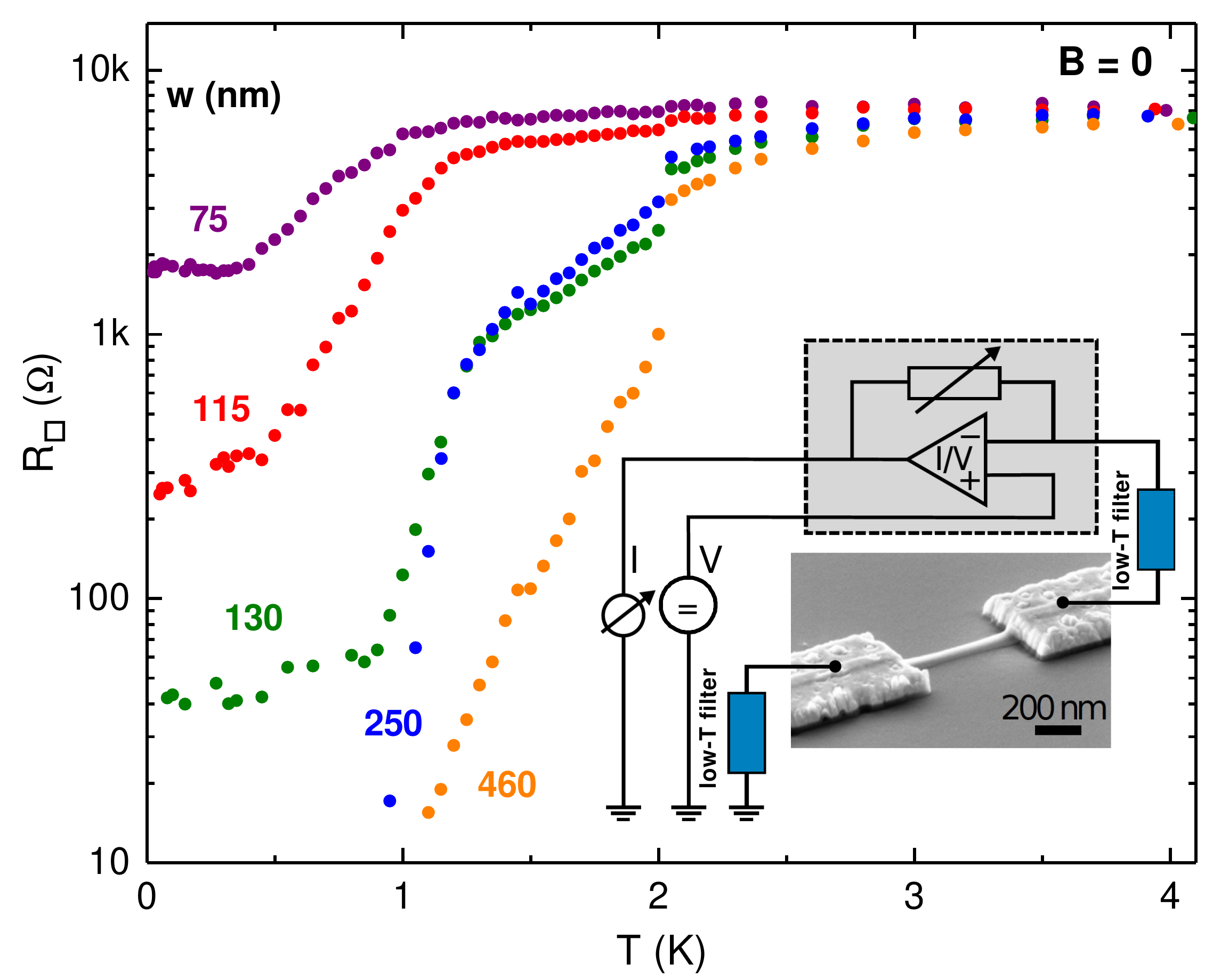}
\caption{\label{fig: RTsc}
Sheet resistance, $R_{\square}$, vs. temperature, $T$, at zero magnetic field of TiN nanostrips of length 780\,nm and with different widths, $w$.   
Inset: scanning electron micrograph of a typical sample and sketch of the measurement setup.
}
\end{figure}

{\it Samples and measurement circuitry}: 
The 780\,nm long nanostrips are prepared by subtractive patterning from a TiN film with a thickness of 3.6\,nm and an initial sheet resistance of 2.5\,k$\Omega$ at 300\,K \cite{Postolova2014}. Across two deposited Au/Ti electrodes with a thickness of $120/10\,$nm, a strip-shaped area of PMMA is exposed to an electron beam at a very high dose (30mC/cm$^2$) that creates cross-links between the polymer strands and renders the exposed areas insoluble. The non-crosslinked PMMA around these areas is then removed in an acetone bath. In a last step the TiN strip is defined by argon ion etching. After processing the sheet resistance $R_\Box$ of the wider strips increased to 4.3\,k$\Omega$ at 300\,K. Because of the deformation of the cross-linked resist during etching the  width of the narrow strips was hard to measure accurately. Most consistent results for the width were obtained from the ratio $R/R_\Box$ at 300\,K. For a design width of 100\,nm the width estimated from $R/R_\Box$ deviated typically by $20\%$. 

We voltage bias the sample to measure the dc-current  with a Femto transimpedance amplifier DDPCA-300-S with integrated biasing circuit. Pi-filters at room temperature and two stages of copper powder filters at mK temperatures provide an efficient high frequency filtering. In some measurements we have added two 50\,k$\Omega$ resistors at the chip carrier in series. The bias resistors shift the load line of the device and provide access also to the usually hysteretic part of the $I$-$V$-characteristics. All resistance values were extracted from the linear regime of the full $I$-$V$-characteristics (see Supplemental Material \cite{supplement}).

{\it Superconducting state}: Figure~\ref{fig: RTsc} shows the linear resistance of several strips vs. temperature $T$. While $R(T)$ approaches zero at low temperatures in the widest strips it clearly saturates for the narrower ones. Such behavior is in line with earlier observations \cite{Bezryadin2008,Lehtinen2017}. The saturation can be explained by a quantum saturation of the phase slip rate at the lowest temperatures. At higher temperatures $R(T)$ can be reasonably well fitted with a simple model of thermally activated phase slips  \cite{supplement}.

\begin{figure}[t]
\includegraphics[width=86mm]{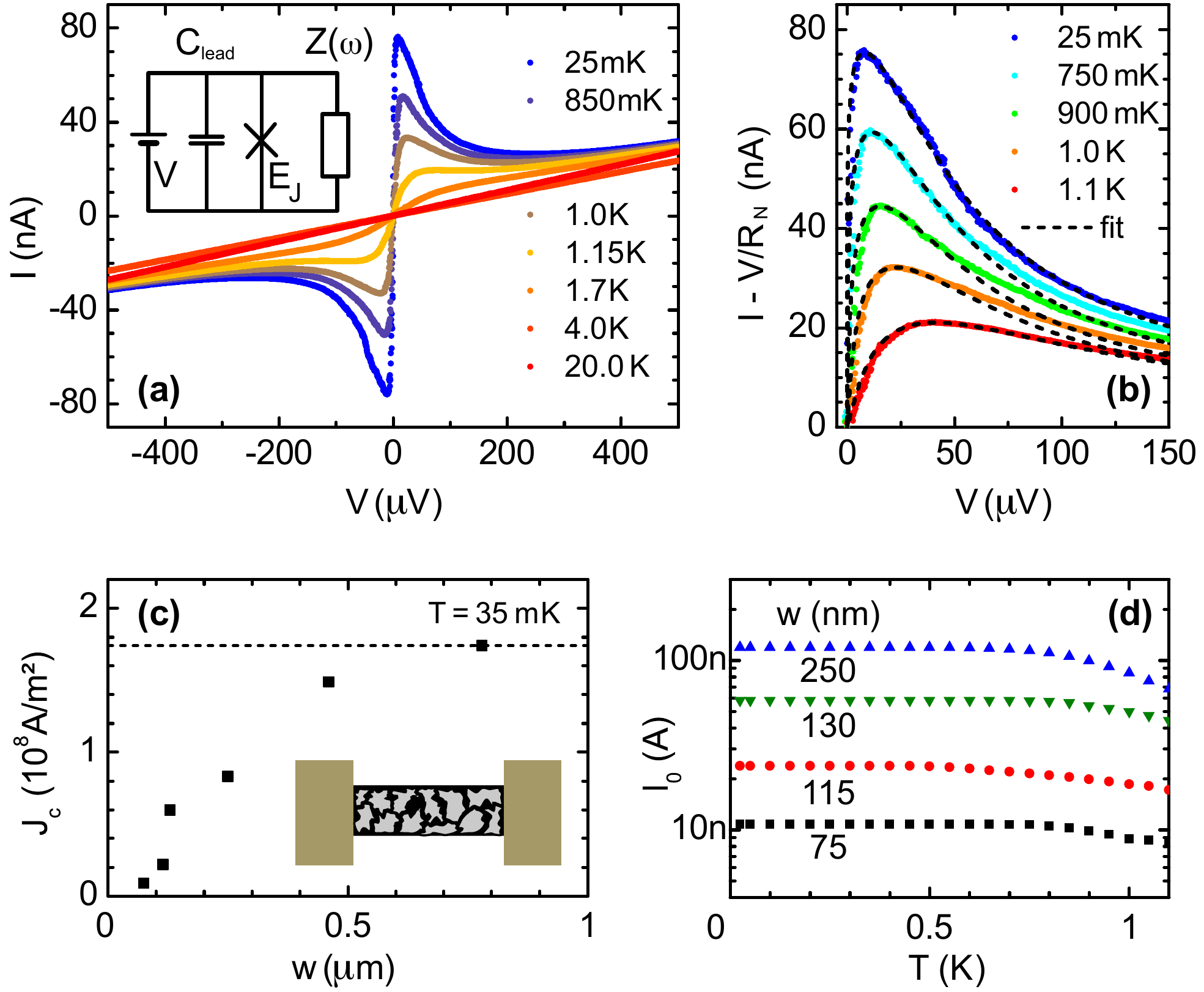}
\caption{\label{fig: lVsc}
(a) Temperature evolution of $I$-$V$ characteristic of a 250\,nm wide strip.  Inset: Voltage source and parallel connection of the active phase slip element and the environmental impedance $Z(\omega)$. The lead capacitance renders the parallel and the serial connection equivalent at high frequencies. (b) Comparison of the data in (a) to the IZ-theory (dashed lines) with $V/R_N$  subtracted from the measured current. (c) Critical current density $J_c$ extracted from the maximal supercurrent vs. nanostrip width. The dashed line indicates the expectation for a standard superconductor with $J_c$ independent of the strip width. Inset: Schematic of droplet structure of the superconducting order parameter. (d) Critical current $I_0$ vs. $T$ for $w=250, \;130 \;115$, and 75\,nm (from top to bottom) as extracted from the IZ-fits (see text). 
}
\end{figure}

The corresponding $I$-$V$ characteristics of the narrower strips display a behavior similar to that of small tunneling junctions \cite{Steinbach2001} with the difference that the current at higher voltages is gradually taken over by a finite parallel resistor with a value $R_N$ that is close to the normal state resistance of the strips. Such behavior is possible, if the transmission probabilities $\tau$ between condensate droplets display a broad distribution between zero and one. If $\tau\gtrsim 0.1$  quasiparticle transport occurs not only by tunneling, but also via Andreev reflection \cite{chauvin}. Within the RCSJ-model, quasiparticle transport can be described a parallel impedance $Z(\omega)$ that is not contained in the IZ-formula (see \cite{supplement} for more details). As an example, the $I$-$V$-characteristics for $w=250$\,nm
is presented in Fig.~\ref{fig: lVsc}a, with $R_N=\mathfrak{Re}[Z(\omega=0)]=20.6\,$k$\Omega$ at 5\,K. Similar observations are made for the  even narrower strips, where the supercurrent peak is gradually suppressed. In contrast, the  $I$-$V$-characteristics of the wider strips display a much more complicated behavior with several different non-equilibrium regimes \cite{supplement}, also observed in large two-dimensional films \cite{ina_tobepubl}.

 If the normal current $V/R_N$ is subtracted  (see Fig.~\ref{fig: lVsc}b for $w=250\,$nm and \cite{supplement} for lower $w$) from the measured  $I$-$V$-characteristics, our data for narrower strips can be well described by the Ivanchenko-Zilberman (IZ) theory for tunnel junctions \cite{Zilberman1969, ambegaokar}. 
 The critical current density $J_c$ extracted from the fitted $I$-$V$-characteristics is plotted vs.~$w$ in Fig.~\ref{fig: lVsc}c. As opposed to a standard superconductor with a width-independent  $J_c$ (dashed line), it dramatically decreases with $w$ indicating that the supercurrent is strongly impeded by phase fluctuations. 
 A very similar overproportional reduction of the switching current $I_\text{sw}/I_\text{sw,max}$ was observed in chains of small Josephson junctions with decreasing ratio $E_J/E_C$ \cite{Pop2010}. In Fig.~\ref{fig: lVsc}d we display the intrinsic critical current  $I_0$ resulting from the fits.  $I_0$ decreases above 600\,mK, consistent with a decrease  of the energy gap $\Delta$ above $T_C/2$. 
 The details of the fitting procedure, a discussion of the fitting parameters, and the $I$-$V$-characteristics for all nanostrips are presented in the supplement \cite{supplement}.

\begin{figure}[t]
\includegraphics[width=1\linewidth]{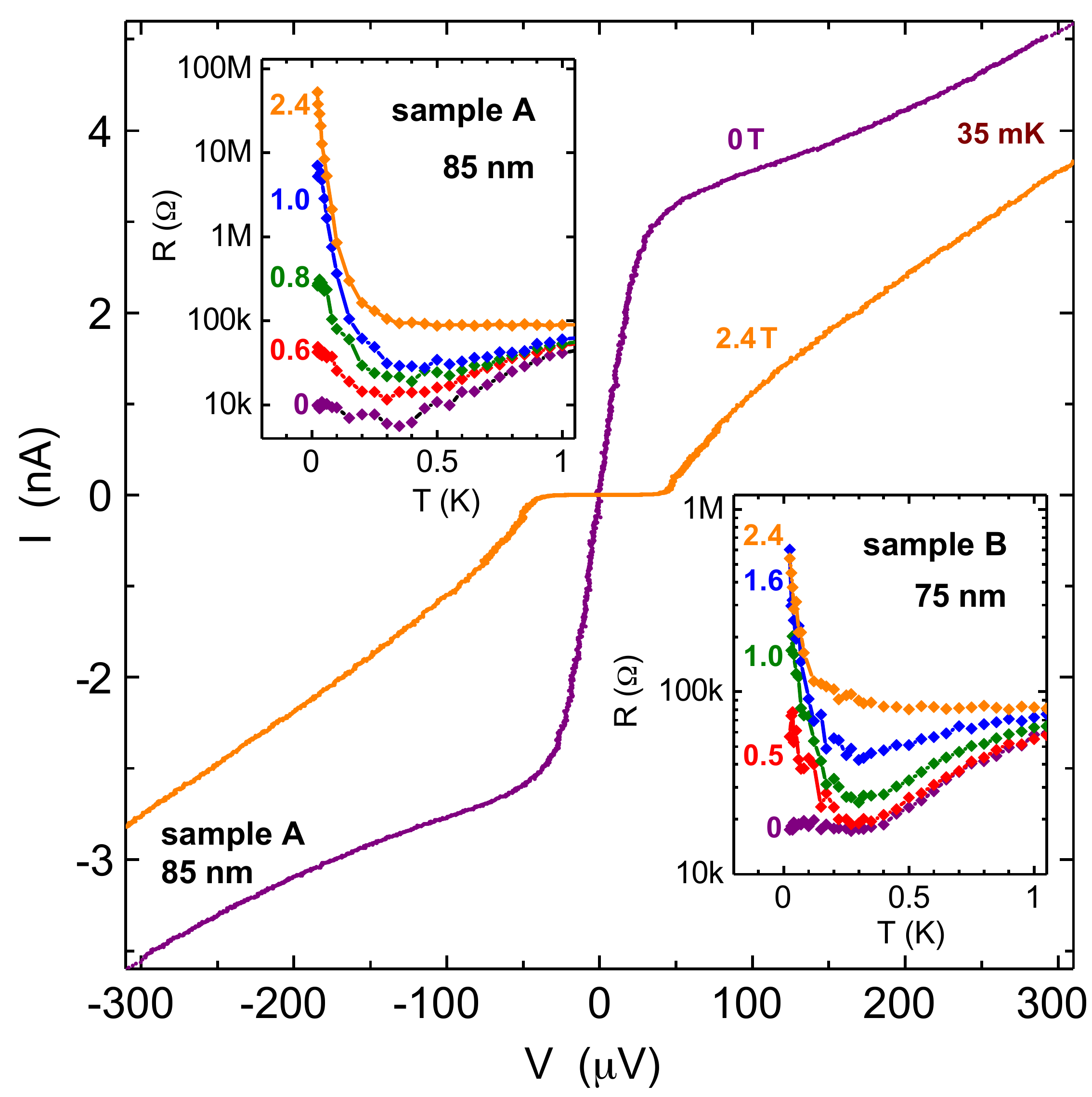}
\caption{\label{fig:RTins}
$I$-$V$-characteristics of nanostrip~{\sf A} ($w=85$\,nm)  for $B=0$ (purple) and in perpendicular magnetic field $B=2.4\,$T (orange). Two 50\,k$\Omega$-resistors were placed adjacent to the device, in order to access the back-bending part of the $I$-$V$-characteristics. Insets:  $R(T)$ for nanostrips~{\sf A}  and~{\sf B} ($w=75$\,nm) with an inititially superconducting and reentrant behavior with increasing magnetic field and for nanostrip~{\sf A}  very high values of the linear resistance.
}\end{figure}

{\it Insulating state}: Next we turn to the behavior in magnetic field.   The purple curve  in Fig.~\ref{fig:RTins} shows the $I$-$V$-characteristics of a nanostrip (labeled {\sf A})  with $w=85$\,nm  [$R(4\,$K$)\simeq 72\,$k$\Omega$] at $B=0$. The superconducting transition is rounded, with the maximum of $I$ missing and the linear resistance $R(T\rightarrow0)\simeq 4\,$k$\Omega$ remaining finite, similar to the narrower nanostrips in Fig.~\ref{fig: RTsc}.  This Josephson-like $I$-$V$-characteristics at $B=0$  can be converted into a strongly insulating one in a perpendicular magnetic field $B=2.4$\,T (orange curve). The linear resistance $R(T)$ of the device is plotted  for different magnetic fields in the upper left inset. In addition, we show a similar evolution of $R(T)$ of another nanostrip (labeled {\sf B})  prepared in a different run with $w=75$\,nm (see Fig.~\ref{fig: RTsc}) in the lower right inset. As seen in both insets to Fig.~\ref{fig:RTins} the resistance initially decreases as superconducting correlations develop in an intermediate temperature regime (0.3\,-\,1\,K). At lower temperatures a reentrant insulating  behavior is observed for both devices.  At the lowest temperatures the linear resistance can exceed the normal state resistance by three orders of magnitude.  

The degree of insulation varies strongly between different devices, and also between different cool-downs of the same device.  Although highly insulating states appear predominantly in the narrower nano\-strips, the behavior is not solely controlled by the nanostrip width, but also depends on  fluctuations of the disorder potential that determines the spatial distribution of the superconducting-order parameter in a sample specific way. 
The  strongly insulating behavior in our narrowest nanostrips ($w<100\,$nm)  in magnetic field is distinct from that of amorphous InO$_\text{x}$ nanowires \cite{Mitra2016}. In the insulating state the resistance of the latter saturates at comparatively low values.

\begin{figure}[t]
\includegraphics[width=86mm]{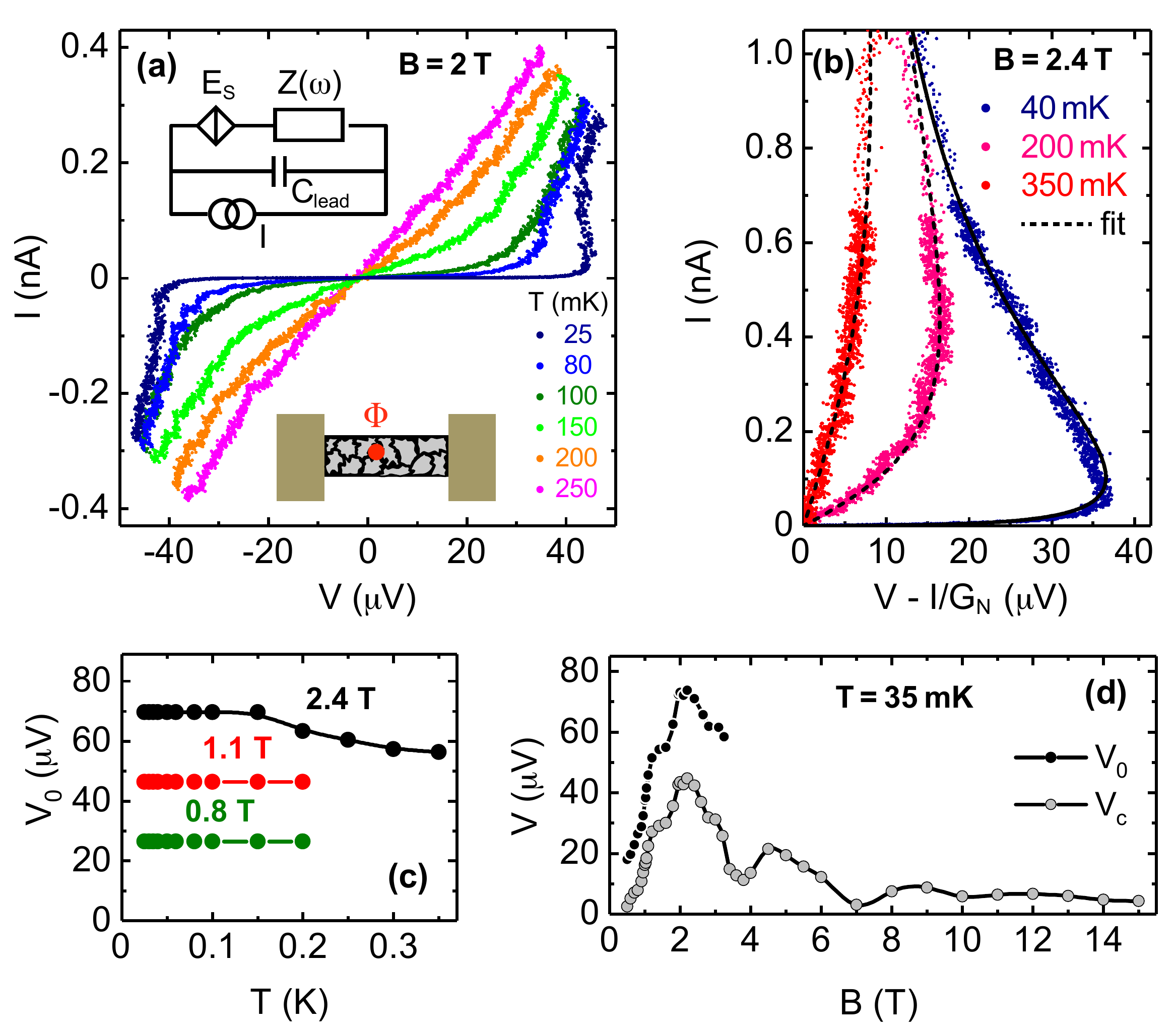}
\caption{\label{fig:IVins}
(a) Evolution of $I$-$V$-characteristics of nanostrip~{\sf A} ($w=85\,$nm) with $T$ at $B=2\,$T in the insulating regime. Upper inset:  Dual measurement circuit with a serial connection of the active phase slip element and the environmental impedance $Z(\omega)$. Lower inset: droplet structure of superconducting order parameter with magnetic flux induced weakest spot (red). (b) Fits with the dual IZ-theory after subtraction of $I/G_N$ from the measured voltage. (c) Temperature dependence of the intrinsic critical voltage $V_0$ extracted from the dual IZ-fits. (d) Magnetic field dependence of the critical voltages $V_c$ and $V_0$.
}\end{figure}

In Fig.~\ref{fig:IVins}a we show the evolution of the $I$-$V$-characteristics of nanostrip~{\sf A} with increasing temperature. A cold bias resistance of 100\,k$\Omega$ allows to access also the back-bending parts of the $I$-$V$-characteristics.  At low $T$ a pronounced Bloch-nose~\cite{likharev1985} develops in the $I$-$V$-characteristics, which is gradually suppressed at higher $T$.  In Figure~\ref{fig:IVins}b we demonstrate that the $I$-$V$-characteristics in the insulating regime are nicely described in terms of the dual Ivanchenko-Zilberman model \cite{corlevi2006}. The voltage drop over the serial impedance $I/G_N$ (which is dual to the contribution $V/R_N$ of a parallel impedance to the current in Fig.~\ref{fig: lVsc}b) is subtracted from the measured bias voltage. The maximal voltages $V_c(T)$ at the Bloch-nose are the dual counterparts of the  current maxima in Fig.~\ref{fig: lVsc}b; $V_c(T)$ is quickly suppressed with increasing temperature.

The corresponding intrinsic critical voltage $V_0$ is extracted from the dual Ivanchenko-Zilberman fit and it exceeds $V_c(T=35\,$mK) by a factor two. It is dual to the intrinsic critical current $I_0$ in Fig.~\ref{fig: lVsc}c,d.  The fitting procedure and the systematic behavior of the fit parameters are described in the supplement \cite{supplement}. The temperature  dependence of $V_0$ is displayed for different magnetic fields  in  Figure~\ref{fig:IVins}c.  In contrast to $V_c(T)$ (not shown) the intrinsic critical voltage $V_0(T)$ remains essentially $T$-independent up to 200~mK (150\,mK for $B=2.4$\,T). 

As a function of magnetic field, $V_c(B)$  plotted in Fig.~\ref{fig:IVins}d displays a strong peak centered around $B\simeq2\,$T that was observed also for the switching voltages in 2d TiN films  \cite{Baturina2007}. Superimposed on the peak  an oscillatory behavior is observed that is reflected also in the magnetoconductance (Fig.~S3 in \cite{supplement}). It is reminiscent to the magnetoconductance fingerprints known from disordered normal metallic wires \cite{LeeStone1985}.  At higher magnetic fields, i.e.~above 3~T (the upper critical field of less disordered TiN-films), it is still possible to simulate the measured $I$-$V$-characteristics using the IZ-model. Because of the fractionalization of the condensate wave function, it is not obvious up to which magnetic field the superconductivity persists locally.


 %

\noindent{\it Discussion}: 
In our devices we use the nanostrip width and the magnetic field to tune the phase slip rate $E_S$  over a wide range, in order to span the  whole transition from a fully superconducting  towards a fully Coulomb blockaded nanostrip. The pronounced sensitivity to the magnetic field can be naturally explained in terms of emerging granularity \cite{Ovadyahu1994,Ghosal2001,Ghosal2001b,Baturina2007,Sacepe2008,Valles2009, Lemarie2013,Carbillet2016}, because the local Josephson coupling energies are gradually suppressed by the magnetic flux enclosed between the granules. Another source for the sensitivity to magnetic field is the global suppression of the local order parameter by the field that is expected to induce a parity effect favoring single electron over Cooper pair tunneling above a certain magnetic field \cite{PhysRevB.91.184505}. 
In zero magnetic field and $w=250\,$nm we find Josephson behavior with a resistance below our detection limit (Fig.~\ref{fig: RTsc}). Reducing $w$ to $\simeq 75\,$nm induces noticable phase fluctuations with an $E_S$ that is sufficiently small to induce a saturation of $R(T)$  (Fig.~\ref{fig: RTsc}).
 The phase fluctuations are manifested further in a gradual suppression of the supercurrent peak (Fig.~\ref{fig:RTins}).  The estimated critical current is ten times smaller than expected from the reduction of the nano\-strip width. 

 In presence of random granularity, the behavior of the strip will be dominated by the weakest spot, where Josephson coupling between the droplets of  the fractionalized condensate is smallest. As shown in Ref.~\onlinecite{Vanevic2012},  rather small variations of the normal state resistance along nominally homogeneously disordered nanowires can induce substantial variations of local quantum phase slip amplitude $E_S$. Differences between different wires are most pronouced in magnetic field. With increasing magnetic field  quantum phase fluctuations at the weakest spot will develop into coherent quantum phase slips -- the corresponding increase of $E_S$, leads to the emergence of Coulomb blockade and eventually strongly insulating behavior. Such behavior has been demonstrated in individual Josephson junctions embedded into 1d-Josephson arrays as high impedance leads~\cite{corlevi2006,Ergul2013}.  Furthermore, the reentrant $R(T)$-curves observed in some of our TiN nanostrips (insets in Fig.~\ref{fig:IVins}) are indistinguishable from those obtained earlier on artificial 1d-Josephson arrays~\cite{Chow1998}.  

 From the maximal value of $E_S(B)=2eV_0(B)/2\pi=4e^2/2C_\text{eff}(B)\simeq 13\,\mu e$V \cite{Mooij2006,Rotzinger2015} observed in Fig.~\ref{fig:IVins}d, we find an effective capacitance  $C_\text{eff}\simeq 0.6\,$fF, which is  much larger than the estimated geometric capacitance $C_\text{geo}\simeq 1.5\,$aF \cite{stray_cap}.  
 The  reduction of $E_S(B)$ with respect to  $4e^2/2C_\text{geo}$ by the residual Josephson coupling is dual to the suppression of $I_0(w)$ (Fig.~\ref{fig: lVsc}c) on the superconducting side.
The effective capacitance has contributions both from the geometric capacitance $C_\text{geo}$ between the granules, and the kinetic capacitance $C_\text{kin}$ that arises from the hybridization between Coulomb parabolas with adjacent numbers of Cooper-pairs \cite{likharev1985}. $C_\text{kin}$ is determined by the inverse curvature of the 2e-periodic energy-charge relation $E(Q)$ (the dual of the energy-phase relation of conventional Josephson junctions) and diverges at the inflection points of $E(Q)$ similar to the divergence of the band mass in a periodic potential. 
In the insulating regime of our critically disordered superconducting nanostrips we expect the kinetic capacitance to dominate over the geometric capacitance in a similar way, as the kinetic inductance dominates over the geometric inductance of our nanostrips at $B=0$.

In conclusion, a magnetic field induced superconduc\-tor-insulator transition is observed in mesoscopic TiN nanostrips. The $I$-$V$-characteristics on both sides of the transition are well described by dual versions of the Ivanchenko-Zilberman model. 
We provide strong evidence that the insulating state is related to Coulomb blockade induced by coherent quantum phase slips. Our observations are consistently explained in terms of a fractionalization of the superconducting condensate in nanostrips made of critically disordered TiN films.\\[2mm]

\begin{acknowledgments} We thank M.~Baklanov for the supply of TiN material, D. Haviland, Yu. Nazarov, A. Ustinov, A. Shnirman and A. Mirlin for inspiring discussions. T.I.B.  acknowledges support by the Alexander von Humboldt Foundation,  by the Ministry of Education and Science of the Russian Federation, and from the Consejer\'{i}a de Educación, Cultura y Deporte (Comunidad de Madrid) through the talent attraction program, Ref.~2016-T3/IND-1839.
\end{acknowledgments}


\begin{thebibliography}{37}
\bibitem [{foo()}]{footnote}%
  \BibitemOpen
  \href@noop {} {\bibinfo  {journal} {Josephson coupling across weak links is
  not restricted to standard tunnel junctions. The presence of highly
  transmissve channels in the weak link typically leads to higher harmonics in
  the current-phase-relation}\ }\BibitemShut {NoStop}%
\bibitem [{str()}]{stray_cap}%
  \BibitemOpen
  \href@noop {} {}\bibinfo {howpublished} {As a rough estimate, we used
  $C_\mathrm{geo}\simeq 8\epsilon_0a$, the capacitance between infinity and a
  disk of radius $a\simeq 20\,$nm. The capacitance between neighboring islands
  is usually smaller.}\BibitemShut {Stop}%
\end{thebibliography}
%

\end{document}